\documentclass[conference]{IEEEtran}
\IEEEoverridecommandlockouts

\pdfoutput=1

\usepackage[cmex10]{amsmath}
\usepackage{cite}

\usepackage{latexsym}
\usepackage{graphics}
\usepackage{amssymb}
\usepackage{amsthm}

\usepackage{cite}
\usepackage{array}
\usepackage{mdwmath}
\usepackage{mdwtab}
\usepackage[tight,footnotesize]{subfigure}
\usepackage{graphicx}

\usepackage{stfloats}
\fnbelowfloat
\usepackage{url}

\ifCLASSOPTIONcompsoc
\usepackage[tight,normalsize,sf,SF]{subfigure}
\else
\usepackage[tight,footnotesize]{subfigure}
\fi

\newcommand{\vect}[1]{\mathbf{#1}}
\def\tr{\mathrm{tr}}
\newcommand{\fracSum}[1]{{\underset{{#1}}{\sum}}}
\newcommand{\maximize}[1]{{\underset{{#1}}{\mathrm{maximize}}}}
\newcommand{\minimize}[1]{{\underset{{#1}}{\mathrm{minimize}}}}

\theoremstyle{remark}

\newtheorem{theorem}{Theorem}
\newtheorem{corollary}{Corollary}

\newtheorem{definition}{Definition}
\newtheorem{example}{Example}

\begin{document}

\title{Optimal Coordinated Beamforming in the \\ Multicell Downlink with Transceiver Impairments}

\author{\IEEEauthorblockN{Emil Bj\"ornson, Per Zetterberg, and Mats Bengtsson}
\IEEEauthorblockA{Signal Processing Lab, ACCESS Linnaeus Center, KTH Royal Institute of Technology (Email: emil.bjornson@ee.kth.se)}
\thanks{The research
leading to these results has received funding from the European
Research Council under the European Community's Seventh Framework
Programme (FP7/2007-2013) / ERC Grant Agreement No. 228044.}}

\maketitle

\begin{abstract}
Physical wireless transceivers suffer from a variety of impairments that distort the transmitted and received signals. Their degrading impact is particularly evident in modern systems
with multiuser transmission, high transmit power, and low-cost devices, but their existence is routinely ignored in the optimization literature for multicell transmission. This paper provides a detailed analysis of coordinated beamforming in the multicell downlink. We solve two optimization problems under a transceiver impairment model and derive the structure of the optimal solutions.
We show numerically that these solutions greatly reduce the impact of impairments, compared with beamforming developed for ideal transceivers.
Although the so-called multiplexing gain is zero under transceiver impairments, we show that the gain of multiplexing can be large at practical SNRs.
\end{abstract}

\IEEEpeerreviewmaketitle

\section{Introduction}

Conventional cellular systems have \emph{fixed} spatial reuse patterns of spectral resources (e.g., time and frequency subcarriers), but modern multi-antenna beamforming enables \emph{dynamic} reuse by exploiting instantaneous channel state information (CSI) \cite{Gesbert2010a}. Under ideal conditions, the downlink can achieve tremendous performance improvements through coordinated multi-antenna transmission among base stations (i.e., cooperative scheduling and beamforming  \cite{Bjornson2011a}). However, the performance of practical cellular systems is limited by various non-idealities, such as computational complexity, CSI uncertainty, limited backhaul capacity, and transceiver impairments.

\begin{figure}[t!]
\includegraphics[width=\columnwidth]{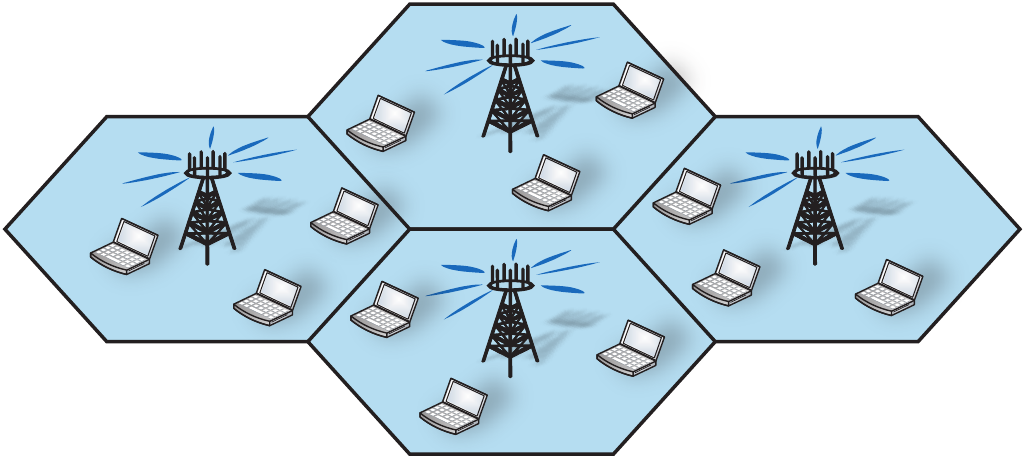} \vskip -3mm
\caption{A multicell system with $N=4$ cells and $K=3$ users per cell. Users are served by their own base station using coordinated beamforming.}\label{figure_cellularsystem} \vskip -4mm
\end{figure}

This paper considers a multicell scenario in which each base station only transmits to its own users, while the beamforming is coordinated among all cells to optimize system performance \cite{Dahrouj2010a}; see Fig.~\ref{figure_cellularsystem}. This setup is known as \emph{coordinated beamforming} and is much easier to implement than the ideal joint transmission case where all base stations serve all users \cite{Bjornson2011a}. Finding the optimal coordinated beamforming is NP-hard under most system performance criteria \cite{Bjornson2012a}, meaning that only suboptimal approaches are feasible in practice. Herein, we concentrate on two system performance criteria that stand out in terms of being globally solvable in an efficient manner:

\begin{list}{$\bullet$}{
\setlength{\leftmargin}{3.5em}
\setlength{\itemindent}{-2em}
}
\item[(P1):] Satisfy quality-of-service (QoS) constraints for each user with minimal power usage;
\item[(P2):] Maximize system performance under some fairness-profile (e.g., maximize worst-user performance).
\end{list}

Both problems can be solved using convex optimization tools or fixed-point iterations; see the seminal works \cite{Rashid1998a,Bengtsson2001a,Wiesel2006a} and recent extensions in \cite{Dahrouj2010a,Bjornson2011a,Bjornson2012a}. These algorithms are usually based on perfect CSI, unlimited backhaul capacity, and ideal transceiver hardware. Recently, the assumption of perfect CSI was relaxed with retained polynomial complexity \cite{Bjornson2012a} and distributed implementation was proposed under certain conditions \cite{Dahrouj2010a,Tolli2009c,Bjornson2013b}. However, ideal transceiver hardware is routinely assumed in the beamforming optimization literature.

Physical hardware implementations of radio frequency (RF) transceivers suffer from impairments such as nonlinear amplifiers, carrier-frequency and sampling-rate offsets, IQ-imbalance, phase noise, and quantization noise \cite{Holma2011a}. The influence of these impairments can be reduced by calibration and compensation algorithms, 
and the residual distortion is often well-modeled by additive Gaussian noise with a power that increases with the power of the useful signal \cite{Dardari2000a,Studer2011a,Studer2010a,Galiotto2009a}.

Transceiver impairments have a relatively minor impact on single-user transmission with low spectral efficiency (e.g., using QPSK \cite{Galiotto2009a}).
The degradations are however particularly severe in modern deployments with small cells, high spectral efficiency, multiuser transmission to low-cost receivers, and transmit-side interference mitigation \cite{Studer2011a}. Still, the existence of impairments is commonly ignored in the development of coordinated multicell schemes, and the optimal scheme is unknown. Prior work has studied point-to-point transmission \cite{Studer2011a,Studer2010a,Galiotto2009a}, non-linear single-cell transmission \cite{Gonzalez2011b}, and multicell zero-forcing transmission \cite{Zetterberg2011a}.
In this paper, we solve \eqref{eq_quality_service_opt} and \eqref{eq_fairness_profile_opt} under a quite general transceiver impairment model and we derive an optimal coordinated beamforming structure.
Numerical examples show how the level of impairments affects performance and that degradations are greatly reduced by taking their existence into account, as done in \eqref{eq_quality_service_opt} and \eqref{eq_fairness_profile_opt}.
A large finite-SNR multiplexing gain can be achieved, although the classic asymptotic multiplexing gain is zero.

\section{System Model}
\label{section_system_model}

We consider the downlink of a multicell system with $N$ cells and $K$ users per cell. Each base station has $N_t$ antennas, while each user has a single antenna. This scenario is illustrated in Fig.~\ref{figure_cellularsystem} and is similar to \cite{Dahrouj2010a}, but we extend the system model in \cite{Dahrouj2010a} by including transceiver impairments. Narrowband subchannels are generated using, for example, \emph{orthogonal frequency-division multiplexing} (OFDM). This paper considers a single subchannel for brevity, but the results are readily extended by adding up the power on all subcarriers in the power constraints and in the characterizations of impairments; see \cite{Bjornson2013b}. The received signal at the $j$th user in the $i$th cell is
\begin{equation}
y_{i,j} = \sum_{m=1}^{N} \vect{h}_{m,i,j}^H \left(\sum_{k=1}^{K} \vect{w}_{m,k} x_{m,k}  +  \vect{z}^{(t)}_{m} \right)+ z^{(r)}_{i,j}.
\end{equation}
The channel vector from the $m$th base station to the $j$th user in the $i$th cell is $\vect{h}_{m,i,j} \in \mathbb{C}^{N_t \times 1}$ and is assumed perfectly known at both sides (to concentrate on other system aspects). The scalar-coded data symbol to this user is circular-symmetric complex Gaussian as $x_{i,j} \sim \mathcal{CN}(0,1)$ and is transmitted using the beamforming vector $\vect{w}_{i,j} \in \mathbb{C}^{N_t \times 1}$. For notational convenience, $\vect{W}_i = [\vect{w}_{i,1} \,\ldots\,  \vect{w}_{i,K}] \in \mathbb{C}^{N_t \times K}$ denotes the combined beamforming matrix in the $i$th cell.

\subsection{Distortions from Transceiver Impairments}
\label{subsection_tx-rx-noise}

The transmission in the $m$th cell is distorted by a variety of transceiver impairments, particularly nonlinear power amplifiers, phase noise, and IQ-imbalance \cite{Galiotto2009a}. After calibrations and compensations, the residual impairments in the transmitter give rise to the additive \emph{transmitter-distortion} term $\vect{z}^{(t)}_{m} \in \mathbb{C}^{N_t \times 1}$. This term is well-modeled as circular-symmetric complex Gaussian because it is the combined residual of many impairments, whereof some are Gaussian and some behave as Gaussian when summed up \cite{Dardari2000a,Studer2010a,Studer2011a,Holma2011a}. The distortion power at a transmit antenna increases with the signal power allocated to this antenna, meaning that $\vect{z}^{(t)}_{m} \sim \mathcal{CN}(\vect{0},\vect{C}_{m})$ where\footnote{Uncorrelated inter-antenna distortion is assumed herein and was validated in \cite{Studer2010a} for transmissions without precoding. We use this reasonable model also with precoding due to the lack of contradicting evidence.}
\begin{equation}
\vect{C}_{m} = \left[\begin{IEEEeqnarraybox*}[][c]{ccc}
c^2_{m,1} & & \\ [-2mm]
& \ddots & \\ [-2mm]
& & c^2_{m,N_t}%
\end{IEEEeqnarraybox*} \right], \quad c_{m,n} = \eta \left( \| \vect{T}_n \vect{W}_{m} \|_F \right).
\end{equation}
The square matrix $\vect{T}_n$ picks out the transmit magnitude at the $n$th antenna (i.e., the $n$th diagonal-element of $\vect{T}_n$ is one, while all other elements are zero). The \emph{monotonically increasing} continuous function $\eta(\cdot)$ of the transmit magnitude (in unit $\sqrt{\text{mW}}$)
models the characteristics of the impairments. These characteristics are measured in the RF-literature using the \emph{error vector magnitude} (EVM) \cite{Studer2010a,Holma2011a}, defined as
\begin{equation}
\mathrm{EVM}_{m,n} \!=\! \frac{\mathbb{E}\big\{ \big|  [\vect{z}^{(t)}_{m}]_{n}  \big|^2 \big\} }{\mathbb{E}\big\{ \big| [\sum_{k} \! \vect{w}_{m,k} x_{m,k} ]_{n}  \big|^2 \big\}} \!=\! \left( \frac{ \eta \left(  \| \vect{T}_n \vect{W}_{m} \|_F  \right) }{ \| \vect{T}_n \vect{W}_{m} \|_F } \right)^2
\end{equation}
for the $n$th transmit antenna in the $m$th cell. $[\cdot]_{n}$ denotes the $n$th element of a vector.
The EVM is the ratio between the average distortion power and the average transmit power, and is often reported in percentage:
$\mathrm{EVM}_{\%}=100 \sqrt{\mathrm{EVM}}$. The EVM requirements for the transmitter in 3GPP Long Term Evolution (LTE) are $8-17.5 \%$, depending on the anticipated spectral efficiency \cite[Section 14.3.4]{Holma2011a}.

\begin{figure}[t!]
\includegraphics[width=\columnwidth]{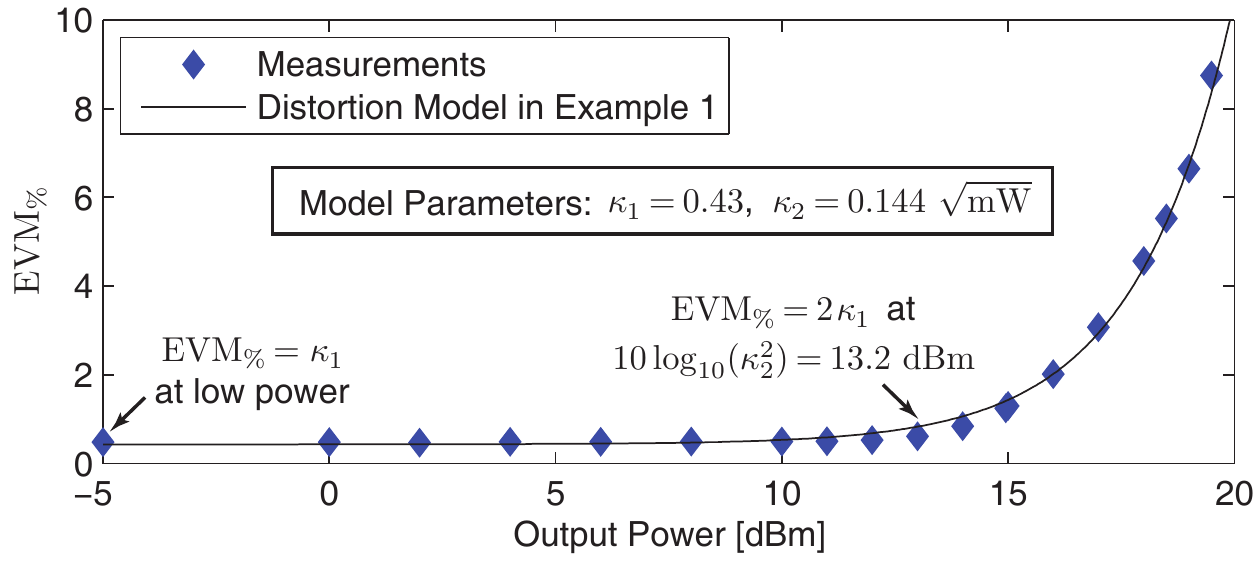} \vskip -2mm
\caption{EVM vs. output power for the LTE power amplifier HXG-122+ in \cite{MinicircuitsHXG-122+} using 64-QAM waveforms and a state-of-the-art signal generator.}\label{figure_evm-model} \vskip -2mm
\end{figure}

\begin{example} \label{example_EVM_tx}
The transmitter-distortion can be modeled as
\begin{equation} \label{eq_TXnoise_example}
\eta(x) = \frac{\kappa_{1}}{100} x\left( 1 + \Big(\frac{x}{\kappa_2}\Big)^4 \right) \quad [\sqrt{\textrm{mW}}]
\end{equation}
where $x = \| \vect{T}_n \vect{W}_{m} \|_F$ is the transmit magnitude and $\kappa_1,\kappa_2$ are model parameters. The first term describes impairments with a constant $\mathrm{EVM}_{\%}$ of $\kappa_{1}$ (e.g., phase noise). The second term models a fifth-order non-linearity in the power amplifier, making $\mathrm{EVM}_{\%}$ increase with $x$; $\mathrm{EVM}_{\%}$ is doubled at  $x=\kappa_2$ $\mathrm{[\sqrt{mW}]}$ and continues to grow. The distortion from the LTE transmitter in \cite{MinicircuitsHXG-122+} is accurately modeled by \eqref{eq_TXnoise_example}; see Fig.~\ref{figure_evm-model}. From the EVM requirements above, $\kappa_{1} \in [0,15]$ is a sensible parameter range.
The designated operating range of the power amplifier is basically upper limited by $10 \log_{10}(\kappa_{2}^2)$ $\mathrm{[dBm]}$.
\end{example}

The reception at the $j$th user in the $i$th cell is distorted by (effective) thermal noise of power $\sigma^2$ and transceiver impairments, particularly phase noise and IQ-imbalance. This is modeled by the complex Gaussian \emph{receiver-distortion} term $z^{(r)}_{i,j} \sim \mathcal{CN}(0,\sigma_{i,j}^2)$ \cite[Section 14.8]{Holma2011a}. The variance is \vskip-3mm
\begin{equation}
\sigma_{i,j}^2 = \sigma^2 +  \nu^2 \left(\sqrt{ \sum_{m=1}^{N}  \| \vect{h}_{m,i,j}^H \vect{W}_{m} \|_F^2 } \right) \quad [\textrm{mW}]
\end{equation}
where $\nu(\cdot)$ models the receiver impairment characteristics and is assumed to be monotonically increasing and continuous.

\begin{example} \label{example_EVM_rx}
The receiver-distortion can be modeled as
\begin{equation} \label{eq_RXnoise_example}
\nu(x) = \frac{\kappa_{3}}{100}  x
\end{equation}
where the model parameter $\kappa_{3} \in [0,15]$ equals $\mathrm{EVM}_{\%}$.
The received signal magnitude $x$ does not change the EVM \cite{Holma2011a}.
\end{example}

\subsection{Transmit Power Constraints}

The transmission in the  $i$th cell is subject to $L_i$ transmit power constraints, which can represent any combination of per-antenna, per-array, and soft-shaping constraints. We write the set of feasible beamforming matrices, $\vect{W}_i$, as \cite{Bjornson2012a}
\begin{eqnarray} \label{eq_power_constraints}
\! \mathcal{W}_i =\! \Big\{ \vect{W}_{i}: \,
\tr( \vect{W}_{i}^H \vect{Q}_{i,k} \vect{W}_{i}) \!+\!  \tr( \delta \vect{Q}_{i,k} \vect{C}_i) \!\leq\! q_{i,k} \,\,\, \forall k \Big\}. \!\!\!
\end{eqnarray}
All $\vect{Q}_{i,k} \in \mathbb{C}^{N_t \times N_t}$ are positive semi-definite matrices and satisfy $\sum_{k=1}^{L_i} \vect{Q}_{i,k} \succ \vect{0}_{N_t} \, \forall i$ (to constrain the power in all spatial directions). For example, per-antenna power constraints of $q$ [mW] are given by $L_i=N_t$, $\vect{Q}_{i,k} = \vect{T}_i$, and $q_{i,k} = q$ for $k=1,\ldots,L_i$. The parameter $\delta \in [0,1]$ determines to what extent the distortions are assumed to consume extra power.

\subsection{User Performance Measure}

The performance of the $j$th user in the $i$th cell is measured by a strictly increasing continuous function $g_{i,j}(\textrm{SINR}_{i,j})$ of the \emph{signal-to-interference-and-noise ratio} (SINR), defined as
\begin{align} \label{eq_SINRij}
&\textrm{SINR}_{i,j}(\vect{W}_{1},\ldots,\vect{W}_{N}) =  \\
&\frac{ |\vect{h}_{i,i,j}^H \vect{w}_{i,j}|^2}{\fracSum{l \neq j} |\vect{h}_{i,i,j}^H \vect{w}_{i,l}|^2 \!+\! \!\fracSum{m \neq i} \! \|\vect{h}_{m,i,j}^H \! \vect{W}_{m}\|_F^2 \!+\! \fracSum{m} \vect{h}_{m,i,j}^H \vect{C}_m \vect{h}_{m,i,j}  \!+\! \sigma_{i,j}^2}. \notag
\end{align}
We let $g_{i,j}(0)=0$ and thus good performance means large positive values on $g_{i,j}(\textrm{SINR}_{i,j})$. This function can, for example, represent the data rate, mean square error, or bit/symbol error rate. The choice of $g_{i,j}(\cdot)$ certainly affects what is the optimal beamforming, but this paper derives optimization algorithms applicable for any choice of performance measures.

\section{Optimization of Coordinated Beamforming}
\label{section_optimal_coordinated_beamforming}

Next, we will compute the optimal coordinated beamforming $\{ \vect{W}_i \}$ under the transceiver impairment model in Section \ref{section_system_model}. We consider two different system performance criteria: \eqref{eq_quality_service_opt} satisfy quality-of-service (QoS) constraints for each user with minimal power usage; and \eqref{eq_fairness_profile_opt} maximize system performance under a fairness-profile. These problems are formulated in this section, efficient solution algorithms are derived, and the solution structure is analyzed.

The first problem is based on having the QoS constraints $g_{i,j}(\textrm{SINR}_{i,j}) \geq \gamma_{i,j}$ for some fixed parameters $\gamma_{i,j}\geq 0$:
\begin{align} \label{eq_quality_service_opt} \tag{P1}
\minimize{\beta,\vect{W}_i \, \forall i} \,\,\,& \,\, \beta \\ \notag
\mathrm{subject \,\,to} \,\, & \,\, g_{i,j}(\textrm{SINR}_{i,j}) \geq \gamma_{i,j} \quad \quad \forall i,j, \\ \notag
& \,\, \tr( \vect{W}_{i}^H \vect{Q}_{i,k} \vect{W}_{i}) +  \tr( \delta \vect{Q}_{i,k} \vect{C}_i) \leq \beta q_{i,k} \quad \forall i,k.
\end{align}
This problem adapts a scaling factor $\beta$ on the power constraints to find the minimal level of power necessary to fulfill all QoS constraints (cf.~\cite{Dahrouj2010a}); thus, an optimal solution to \eqref{eq_quality_service_opt} with $\beta \leq 1$ means that the QoS constraints are feasible under the power constraints in \eqref{eq_power_constraints}.\footnote{If the QoS constraints are too optimistic, the co-user interference and hardware impairments might make it impossible to satisfy all constraints irrespectively of how much power that is used. Thus, \eqref{eq_quality_service_opt} can be infeasible.} Observe that \eqref{eq_quality_service_opt} can also be formulated as a feasibility problem (by fixing $\beta=1$), but this is not necessarily more computationally efficient in practice.

As it might be difficult to find good QoS constraints \emph{a priori}, the second problem includes these parameters in the optimization by replacing them with two fairness constraints:
\begin{itemize}
\item Each user has a predefined lowest acceptable QoS level $a_{i,j} \geq 0$; thus, $g_{i,j}(\textrm{SINR}_{i,j}) \geq a_{i,j}$.
\item Each user gets a predefined portion $\alpha_{i,j} \geq 0$ of the exceeding performance, where $\sum_{i,j} \alpha_{i,j}=1$.
\end{itemize}
The corresponding problem is known as a \emph{fairness-profile optimization} (FPO) problem \cite{Bjornson2012a}:
\begin{equation} \label{eq_fairness_profile_opt} \tag{P2}
\begin{split}
\maximize{\vect{W}_i \in \mathcal{W}_i \, \forall i} \,\,\,& \,\, \min_{i,j} \, \frac{g_{i,j}(\textrm{SINR}_{i,j}) - a_{i,j}}{\alpha_{i,j}} \\
\mathrm{subject \,\,to} \,\, & \,\, g_{i,j}(\textrm{SINR}_{i,j}) \geq a_{i,j} \quad \forall i,j.
\end{split}
\end{equation}
This is a recent generalization of classic max-min optimization (cf.~\cite{Wiesel2006a}) that handles heterogenous user channel conditions by selection of $a_{i,j},\alpha_{i,j}$.
The FPO problem is infeasible if $a_{i,j}$ is too large, thus the system can select them pessimistically to guarantee a minimal QoS level and rely on that \eqref{eq_fairness_profile_opt} optimizes the actual QoS based on the current channel conditions.

\subsection{Convex and Quasi-Convex Reformulations of \eqref{eq_quality_service_opt} and \eqref{eq_fairness_profile_opt}}

Under ideal transceiver hardware, considerable attention has been given to various forms of the optimization problems \eqref{eq_quality_service_opt} and \eqref{eq_fairness_profile_opt}. Efficient solution algorithms have been proposed for both single-cell and multi-cell systems; see \cite{Rashid1998a,Bengtsson2001a,Wiesel2006a,Dahrouj2010a,Bjornson2011a,Bjornson2012a} and reference therein.
Next, we show how these results can be generalized to also include the distortion generated by hardware impairments in the transmitters and receivers. We will first solve \eqref{eq_quality_service_opt} and then show how that solution can be exploited to solve \eqref{eq_fairness_profile_opt} in a simple iterative manner.

\begin{theorem} \label{theorem_QoS}
Let $\gamma_{i,j}\geq 0$ be given. If $\eta(\cdot)$ and $\nu(\cdot)$ are monotonic increasing convex functions, then \eqref{eq_quality_service_opt} can be reformulated into the following convex optimization problem:
\begin{align} \label{eq_quality_service_opt_convex}
&\minimize{\beta,\vect{W}_i, t_{i,\!n}, r_{i,\!j} \, \forall i,j,n} \,\,\, \,\, \beta \\
& \quad \, \mathrm{subject \,\,to} \quad t_{i,n} \!\geq\! 0, \,\, r_{i,j} \!\geq\! 0, \,\, \Im( \vect{h}_{i,i,j}^H \vect{w}_{i,j} )\!=\!0 \quad \! \forall i,j,n, \notag \\
& \!\! \tr( \vect{W}_{i}^H \vect{Q}_{i,k} \vect{W}_{i}) +  \sum_{n} \tr( \delta \vect{Q}_{i,k} \vect{T}_n) t_{i,n}^2 \!\leq\! \beta q_{i,k} \quad \forall i,k, \label{eq_const1} \\
& \! \!\sqrt{\sum_{m} \|\vect{h}_{m,i,j}^H \! \vect{W}_{m}\|_F^2
+ \sum_{m,n} (\vect{h}_{m,i,j}^H \vect{T}_n \vect{h}_{m,i,j}) t_{m,n}^2 \!+\! r_{i,j}^2 \!+\! \sigma^2} \notag \\[-1mm]
& \qquad \qquad \leq \sqrt{ 1+ \frac{1}{g^{-1}_{i,j}(\gamma_{i,j})} } \, \Re ( \vect{h}_{i,i,j}^H \vect{w}_{i,j}) \quad  \forall i,j,  \label{eq_const2}\\
& \qquad \qquad \eta ( \| \vect{T}_n \vect{W}_{m} \|_F ) \leq  t_{m,n} \quad \forall m,n,  \label{eq_const3} \\
& \qquad \qquad \nu \Big(\sqrt{\sum_{m}  \| \vect{h}_{m,i,j}^H \vect{W}_{m} \|_F^2} \Big) \leq r_{i,j} \quad  \forall i,j.  \label{eq_const4}
\end{align}
\end{theorem}
\begin{IEEEproof}
The proof is given in the appendix.
\end{IEEEproof}

The convexity of $\eta(\cdot),\nu(\cdot)$ is a rather reasonable assumption that is satisfied by any polynomial function with positive coefficients (e.g., Examples \ref{example_EVM_tx} and \ref{example_EVM_rx}). It means that the distortion power increases equally fast or faster than the signal power.

Theorem \ref{theorem_QoS} proves that \eqref{eq_quality_service_opt} is a convex problem (under reasonable conditions), meaning that the optimal solution can be obtained in polynomial time (e.g., using general-purpose implementations of interior-point methods \cite{cvx}). The theorem extends previous convexity results for multi-cell systems in \cite{Dahrouj2010a,Bjornson2012a,Bjornson2011a} to also include transceiver impairments. Distributed implementation is possible using a dual decomposition approach with limited backhaul signaling, similar to \cite{Tolli2009c,Bjornson2013b}.

Next, we give a corollary that shows how the FPO problem \eqref{eq_fairness_profile_opt} can be solved efficiently using Theorem \ref{theorem_QoS}.

\begin{corollary} \label{corollary_FPO}
For given $a_{i,j},\alpha_{i,j}$ and an upper bound $f^{\text{upper}}_{\text{FPO}}$ on the optimum of \eqref{eq_fairness_profile_opt}, the problem can be solved by bisection over $\mathcal{F}=[0,f^{\text{upper}}_{\text{FPO}}]$. For given $f_{\text{candidate}} \in \mathcal{F}$, we try to solve \eqref{eq_quality_service_opt} for $\gamma_{i,j}=g_{i,j}^{-1} (a_{i,j} \!+\! \alpha_{i,j} f_{\text{candidate}} )$ using Theorem \ref{theorem_QoS}.

If \eqref{eq_quality_service_opt} is feasible for these $\gamma_{i,j}$ and $\beta_{\text{solution}}\leq 1$, then all $\tilde{f} \in \mathcal{F}$ with $\tilde{f}< f_{\text{candidate}}$ are removed. Otherwise, all $\tilde{f} \in \mathcal{F}$ with $\tilde{f} \geq f_{\text{candidate}}$ are removed.
\end{corollary}
\begin{IEEEproof}
The algorithm searches on a line in the so-called performance region; see further details and proofs in \cite{Bjornson2012a}.
\end{IEEEproof}

This corollary shows that \eqref{eq_fairness_profile_opt} can be solved through a series of QoS problems of the type in \eqref{eq_quality_service_opt}. Since each subproblem is convex and bisection has linear convergence, we conclude that the FPO problem with transceiver impairments is quasi-convex and can be solved in polynomial time \cite{cvx}.

Observe that Corollary \ref{corollary_FPO} requires an initial upper bound $f^{\text{upper}}_{\text{FPO}}$, but it is easy to achieve by relaxing the problem (e.g., by ignoring all interference and impairments); see \cite{Bjornson2012a}.

\begin{figure}
\begin{center}
\includegraphics[width=80mm]{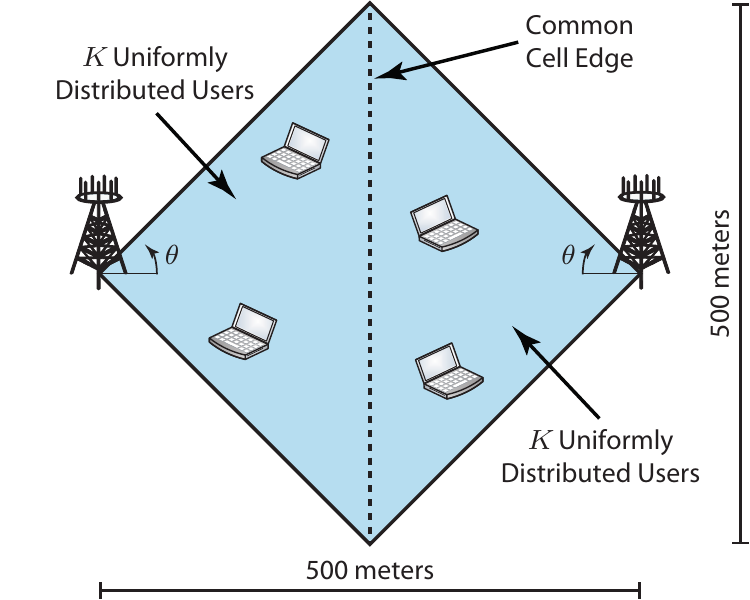}
\end{center} \vskip-4mm
\caption{Illustration of the simulation scenario.}\label{figure_simulation_scenario} \vskip-3mm
\end{figure}

\subsection{Structure of the Optimal Coordinated Beamforming}

Next, we investigate the optimal beamforming structure. The beamforming vectors can be decomposed as $\vect{w}_{i,j} = \sqrt{p_{i,j}} \vect{v}_{i,j}$.

\begin{theorem}
If \eqref{eq_quality_service_opt} or \eqref{eq_fairness_profile_opt} is feasible, it holds that:
\begin{itemize}
\item The optimal beamforming direction $\vect{v}_{i,j}$ is equal to
\begin{equation*} \label{eq_beamforming_structure}
\frac{
\Big( \fracSum{k} \lambda_{i,k} \vect{Q}_{i,k} \!+\! \fracSum{m,l} \mu_{m,l} \vect{h}_{i,m,l} \vect{h}_{i,m,l}^H \!+\! \fracSum{n} \tau_{i,n} \vect{T}_n \! \Big)^{\!\!-1} \vect{h}_{i,i,j}
}{\Big\| \! \Big( \fracSum{k} \lambda_{i,k} \vect{Q}_{i,k} \!+\! \fracSum{m,l} \mu_{m,l} \vect{h}_{i,m,l} \vect{h}_{i,m,l}^H \!+\! \fracSum{n} \tau_{i,n} \vect{T}_n \! \Big)^{\!\!-1} \vect{h}_{i,i,j} \Big\| }
\end{equation*}
for some $[0,1]$-parameters $\{\lambda_{i,k}\}$, $\{\mu_{m,l}\}$, and $\{\tau_{i,n}\}$.

\item The optimal power allocation $p_{i,j}$ is smaller than some fixed $\tilde{q}<\infty$ irrespectively of the power constraints, if $x/\eta(x) \rightarrow 0$ or $ x/\nu(x) \rightarrow 0$ as $x \rightarrow \infty$.
\end{itemize}
\end{theorem}
\begin{IEEEproof}
The beamforming direction structure is achieved by the approach in \cite[Theorem 3]{Bjornson2011a}. If $\eta(\cdot)$ or $\nu(\cdot)$ grow faster than linear, it easy to show that $\textrm{SINR}_{i,j} \rightarrow 0$ as $p_{i,j} \rightarrow \infty$. The optimal power allocation will therefore always be bounded.
\end{IEEEproof}

The first property shows that the beamforming direction has a similar structure as for ideal transceivers (cf.~\cite{Bjornson2011a}). The only difference is that the optimal parameters generally are different and that $\sum_{n} \tau_{i,n} \vect{T}_n$ acts as extra per-antenna constraints.

The second property means that if the distortion power scales faster than the signal power (e.g., as in Example \ref{example_EVM_tx} with $\kappa_2<\infty$), there is an upper limit on how much power that should be used. Being above this limit will only hurt the performance, even in simulation scenarios where $p_{i,j} \rightarrow \infty$ would have meant that
$\textrm{SINR}_{i,j} \rightarrow \infty$ if the transceivers would have been ideal.
The impact of this property on the multiplexing gain is discussed in the next section.

\section{Numerical Evaluation}

Next, we illustrate numerically the impact of transceiver impairments on the throughput of coordinated multicell systems. We consider a simple scenario with $N=2$ base stations located in the opposite corners of a square (with diagonal of 500 meters); see Fig.~\ref{figure_simulation_scenario}. The data rate, $g_{i,j}(\textrm{SINR}_{i,j})=\log_2(1+\textrm{SINR}_{i,j})$, is used as user performance measure and $K$ indoor users are uniformly distributed in each half of the coverage area (at least 35 meters from the base station). The fixed system parameters are summarized in Table \ref{table_system_parameters}. The system is best described as a simplified version of Case 1 in the 3GPP LTE standard \cite{LTE2010b} where we assume uncorrelated Rayleigh fading channels and independent shadowing.

We compare two coordinated beamforming approaches.
\begin{itemize}
\item Max-min optimized beamforming: The solution to \eqref{eq_fairness_profile_opt} with $a_{i,j}=0$, $\alpha_{i,j}=\frac{1}{NK}$, achieved by Corollary \ref{corollary_FPO}.

\item Distortion-ignoring beamforming: The solution to \eqref{eq_fairness_profile_opt} with ideal transceivers (i.e., $\nu(\cdot)\!=\!\eta(\cdot)\!=\!0$) and $a_{i,j}\!=\!0$, $\alpha_{i,j}=\frac{1}{NK}$. Similar to max-min approaches in \cite{Wiesel2006a,Bjornson2012a}.
\end{itemize}

These approaches coincide and maximize the worst-user performance with ideal transceivers, while only the former one is optimal under impairments.
In fact, the latter uses more power than allowed in \eqref{eq_power_constraints}, but this has negligible impact.

\begin{table}[!t]
\renewcommand{\arraystretch}{1.3}
\caption{Fixed System Parameters in the Numerical Evaluation}
\label{table_system_parameters} \vskip-2mm
\centering
\begin{tabular}{|c|c|}
\hline
\bfseries Parameters & \bfseries Values\\
\hline

Transmit Antenna Gain Pattern,  $\theta \in [-\frac{\pi}{4},\frac{\pi}{4}]$& $14 - 8 \, \theta^2 $ dB \\

Receive Antenna Gain & 0 dB \\

Carrier Frequency / Downlink Bandwidth & 2 GHz / 10 MHz\\

Number of Subcarriers /  Bandwidth & 600 / 15 kHz \\

Small Scale Fading Distribution & $\mathcal{CN}(\vect{0},\vect{I})$ \\

Standard Deviation of Lognormal Shadowing & 8 dB\\

Path Loss at Distance $d$ (kilometers) & \!$128.1 \!+\! 37.6 \log_{10}(d)$\! \\

Penetration Loss (indoor users) & 20 dB \\

Noise Power $\sigma^2$ (5 dB noise figure) & $-127$ dBm \\

\hline
\end{tabular} \vskip-4mm
\end{table}

\begin{figure}[t!]
\subfigure[Fixed receiver and varying transmitter-distortion ($\kappa_3=2$).]
{\label{figure_txdist}\includegraphics[width=\columnwidth]{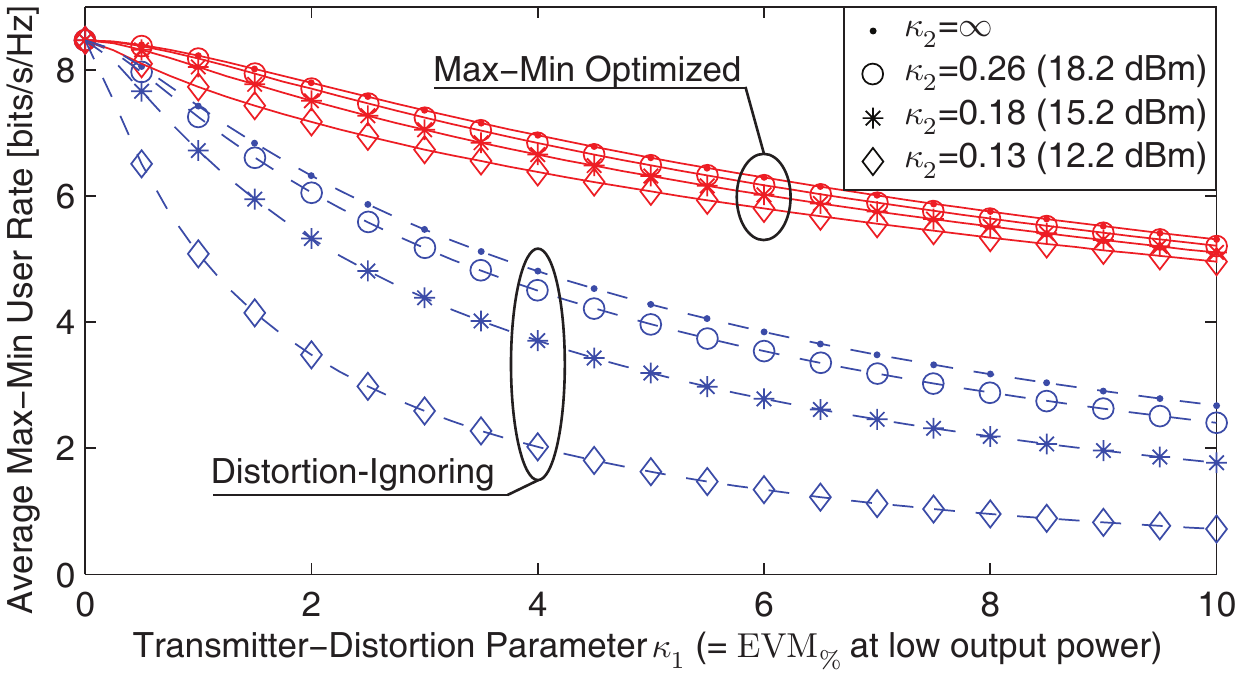}}\hfill
\subfigure[Fixed transmitter and varying receiver-distortion ($\kappa_2=\infty$, different $\kappa_1$).]
{\label{figure_rxdist}\includegraphics[width=\columnwidth]{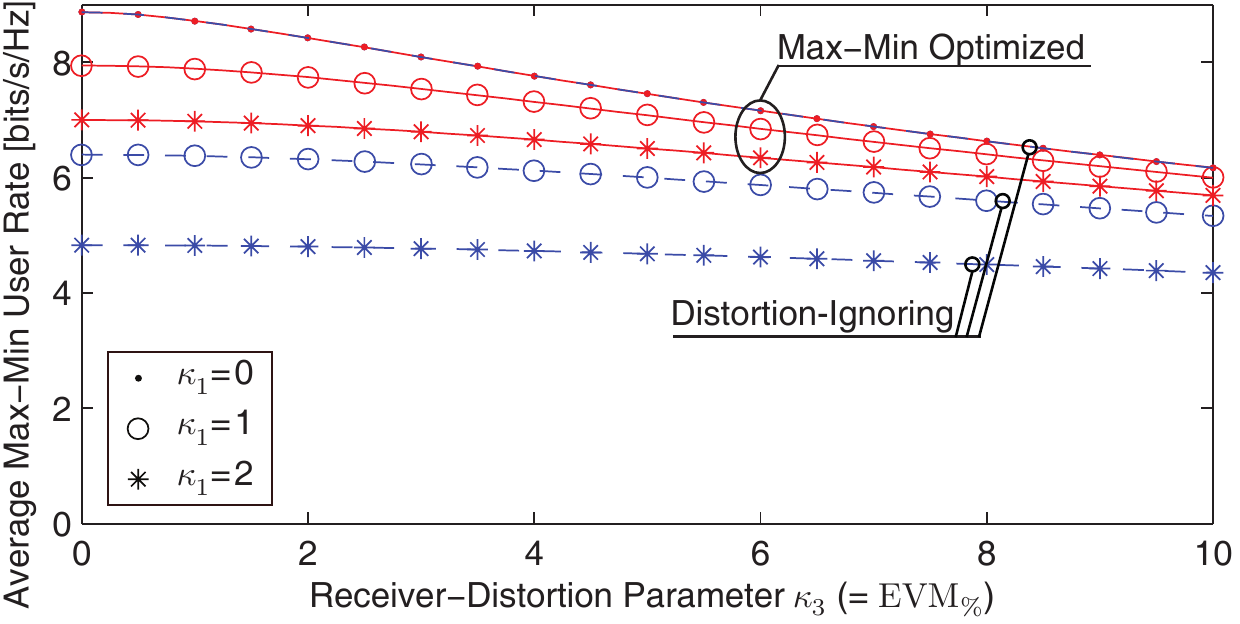}}\hfill
\subfigure[Varying transmitter/receiver-distortion ($\kappa_1=\kappa_3$, different $\kappa_2$).]
{\label{figure_tx-rxdist}\includegraphics[width=\columnwidth]{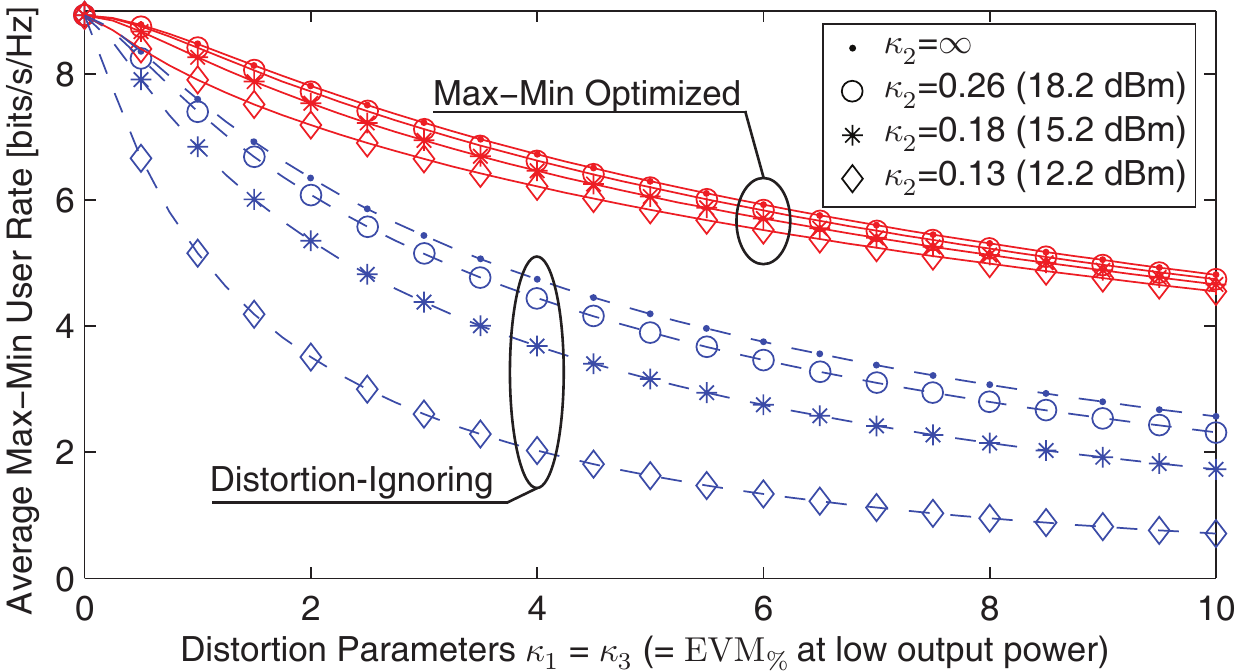}}
\caption{Average max-min user rates with varying transceiver impairments. The optimal coordinated beamforming given by \eqref{eq_fairness_profile_opt} is compared with the corresponding beamforming approach when all impairments are ignored.} \label{figure_distortion} \vskip-4mm
\end{figure}

\subsection{Impact of Transceiver Impairment Characteristics}

First, we consider $K=2$ users per cell, $N_t=4$ transmit antennas, and per-array power constraints of 18.2 dBm per subcarrier (i.e., uniform allocation of 46 dBm). We study how the max-min user rate is affected by the level of transceiver impairments. The impairments are modeled as in Examples \ref{example_EVM_tx} and \ref{example_EVM_rx} (with $\delta=1$) and we will vary the parameters $\kappa_1,\kappa_2,\kappa_3$.

The average user performance (over channel realizations and user locations) is shown in Fig.~\ref{figure_distortion}: (a) considers fixed receiver-distortion ($\kappa_3=2$) and varying transmitter-distortion; (b) considers fixed transmitter-distortion ($\kappa_2=\infty$, different $\kappa_1$) and varying receiver-distortion; and (c) varies both transmitter- and receiver-distortion ($\kappa_1=\kappa_3$, different $\kappa_2$).

The main observation is that transceiver impairments cause substantial performance degradations, unless $\mathrm{EVM}_{\%}<1$.
High-quality hardware is therefore required to operate close to the ideal performance.
The performance loss can however be reduced by taking the impairment characteristics (particularly transmitter-distortion) into account in the beamforming selection. The optimization procedure in Section \ref{section_optimal_coordinated_beamforming} enables higher data rates or the same rates using less expensive transceivers (with 2-9 percentage points larger $\mathrm{EVM}_{\%}$).
Decreasing $\kappa_2$ will reduce the designated operating range of the amplifier (i.e., where the EVM is almost constant). If $\kappa_2^2$ is smaller than the power constraint, the optimal beamforming will use less power than available. Distortion-ignoring beamforming becomes highly suboptimal in these cases as it uses full power.

\subsection{Impact on Multiplexing Gain}

Next, we investigate how coordinated beamforming improves the sum rate compared with time division multiple access (TDMA). This can be characterized using the \emph{multiplexing gain}, defined as the slope of the sum rate versus output power curve in the high-power regime \cite{Gesbert2010a}. Coordinated beamforming can obtain a multiplexing gain of $\min(N_t,NK)$ with ideal transceivers, meaning that the sum rate behaves as $\min(N_t,NK) \log_2(q) + \mathcal{O}(1)$ where $q$ is the output power.

\begin{figure}[t!]
\includegraphics[width=\columnwidth]{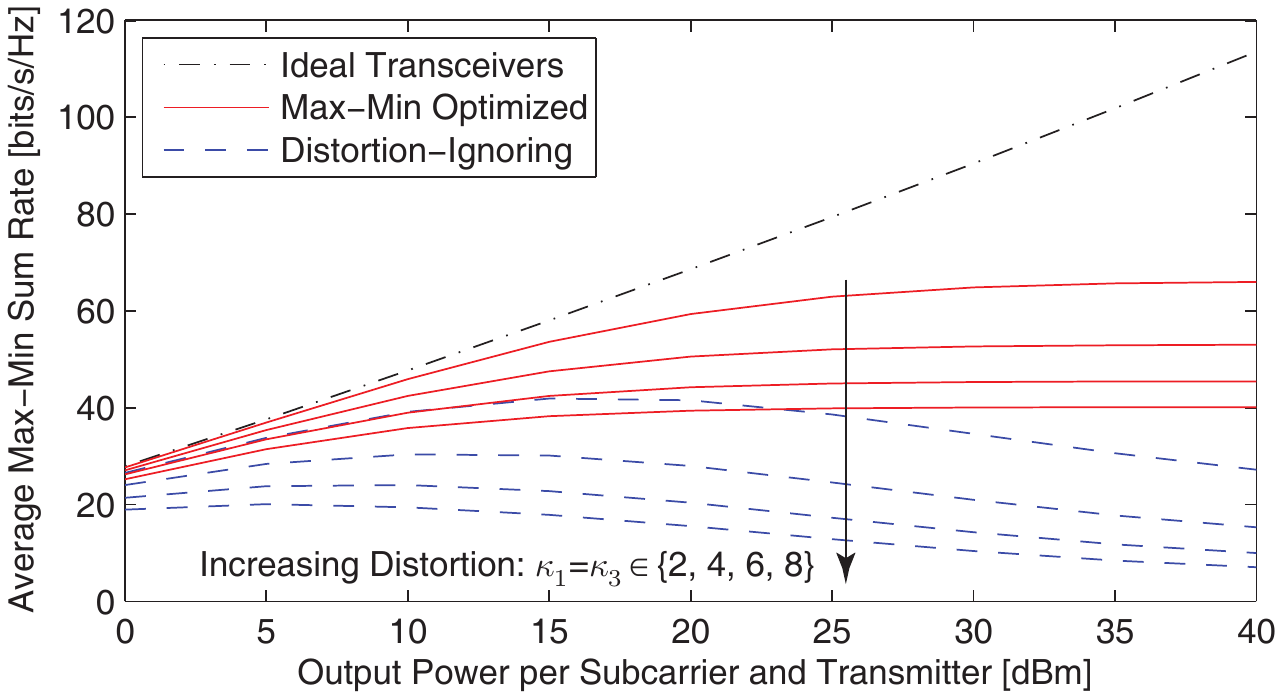} \vskip -2mm
\caption{Average max-min sum rate as a function of the output power and for different levels of transceiver impairments. Optimal coordinated beamforming from \eqref{eq_fairness_profile_opt} is compared with
the counterpart when impairments are ignored.}\label{figure_SNR} \vskip -4mm
\end{figure}

The average max-min sum rate (over channel realizations and user locations) is shown in Fig.~\ref{figure_SNR} as a function of the output power per transmitter and subcarrier. We have $N_t=8$ antennas and $N=4$ users per cell. The sum rate with ideal transceivers is compared with having impairments with $\kappa_1 = \kappa_3 \in \{ 2, 4, 6, 8\}$ and $\kappa_2 = \infty$. We consider both max-min optimized beamforming and distortion-ignoring beamforming.

As expected, the ideal sum rate in Fig.~\ref{figure_SNR} achieves a multiplexing gain of 8. On the other hand, the sum rate with transceiver impairments is bounded and decreases with exacerbated impairments---therefore, only \emph{zero} multiplexing gain is achievable under transceiver impairments, which is natural since the distortion increases with the output power and creates an irreducible error floor. The sum rate with optimal max-min beamforming converges to a rather high level, while distortion-ignoring beamforming behaves strangely; the sum rate actually decreases in the high-power regime because it converges to suboptimal zero-forcing beamforming.

The existence of a multiplexing gain (i.e., sum rate growing unboundedly with the output power) can be viewed as an artifact of ignoring the transceiver impairments that always appear in practice.\footnote{Mathematically, a non-zero multiplexing gain can be achieved if $\mathrm{EVM}_{\%}$ would \emph{decrease towards zero} as the signal power increases, but this is highly unreasonable since the EVM typically \emph{increases} with the signal power.} However, also practical systems can gain from multiplexing and thus we use an alternative definition:

\begin{definition} \label{definition_multiplexing_slope}
The \emph{finite-SNR multiplexing gain} $M_{\rho}$ is the ratio between the average sum rate for a coordinated beamforming strategy and the
average TDMA rate at same output power $\rho$.
\end{definition}

This definition refines the one in \cite{Narasimhan2005b} and makes $M_{\rho}$ the average multiplicative gain of coordinated beamforming over optimal TDMA. If
$M_{\rho} \gg 1$, coordinated beamforming could be useful in practice (where the CSI uncertainty typically decreases $M_{\rho}$).
The finite-SNR multiplexing gain is shown in Fig.~\ref{figure_multiplexinggain} for the same scenario as in Fig.~\ref{figure_SNR}. We observe that $M_{\rho}$
is actually higher under transceiver impairments than with ideal hardware (at practical output power), because the average rate with TDMA saturates under impairments.
Coordinated beamforming is therefore particularly important for boosting performance under impairments. We also note that this advantage is lost if the distortions are ignored.

\section{Conclusion}

Transceiver impairments greatly degrade the performance of coordinated beamforming systems, particularly if their characteristics are ignored in the transmission design. This paper derived the optimal beamforming under transceiver impairments for two system performance criteria: satisfy QoS constraints and fairness-profile optimization. The solutions reduce the performance losses, thus enabling higher throughput or the use of less expensive hardware. We also derived the optimal beamforming structure and showed that there is an upper performance bound, irrespectively of the available power.
Interestingly, impairments can make coordinated beamforming even more favorable than under ideal transceivers, because the finite-SNR multiplexing gain can be larger.

\appendix

\textit{Proof of Theorem \ref{theorem_QoS}:} The cost function of \eqref{eq_quality_service_opt} is convex, thus it remains to achieve convex constraints.
First, we represent the distortion power at the $n$th antenna of the $m$th base station by the new variables $t_{m,n} = c_{m,n} = \eta( \|  \vect{T}_n \vect{W}_{m} \|_F )$.
This makes $\vect{h}_{m,i,j}^H \vect{C}_m \vect{h}_{m,i,j} = \sum_{n} (\vect{h}_{m,i,j}^H \vect{T}_n \vect{h}_{m,i,j}) t_{m,n}^2$ and $\tr( \vect{Q}_{m,k} \vect{C}_m) = \tr( \vect{Q}_{m,k} \vect{T}_n) t_{m,n}^2$ in the QoS and power constraints, respectively. By minimizing over $t_{m,n}$, the equality can be relaxed as \eqref{eq_const3} and is a convex constraints if the increasing function $\eta(\cdot)$ is convex.

Similarly, we introduce $r_{i,j}$ to represent the receiver-distortion power as
 $r^2_{i,j} = \nu^2 (\sqrt{ \sum_m \| \vect{h}_{m,i,j}^H \vect{W}_{m} \|_F^2 } )$. By minimizing over $r_{i,j}$, this relationship can be expressed as \eqref{eq_const4}
 and is convex if the increasing function $\nu(\cdot)$ is convex.

Using the new variables, the power constraints in \eqref{eq_const1} are quadratic and convex. The constraints $g_{i,j}(\textrm{SINR}_{i,j}) \geq \gamma_{i,j}$ can be written as $\textrm{SINR}_{i,j} \geq g^{-1}_{i,j}(\gamma_{i,j})$  and expanded as
\begin{equation} \label{eq_SINRij_new_variables}
\begin{split}
\! \! \frac{|\vect{h}_{i,i,j}^H \vect{w}_{i,j}|^2}{g_{i,j}^{-1}(\gamma_{i,j})}& \geq  \Big(
\fracSum{l \neq j} |\vect{h}_{i,i,j}^H \vect{w}_{i,l}|^2 \!+\! \!\fracSum{m \neq i} \|\vect{h}_{m,i,j}^H \! \vect{W}_{m}\|_F^2 \\
& + \fracSum{m,n} (\vect{h}_{m,i,j}^H \vect{T}_n \vect{h}_{m,i,j}) t_{m,n}^2  + r_{i,j}^2 + \sigma^2 \Big)
\end{split}
\end{equation}
by moving around the terms. Finally, \eqref{eq_SINRij_new_variables} is reformulated as the convex second-order cone in \eqref{eq_const2} by exploiting the phase ambiguity in $|\vect{h}_{i,i,j}^H \vect{w}_{i,j}|$ to set $\vect{h}_{i,i,j}^H \vect{w}_{i,j}>0$ (cf.~\cite{Bengtsson2001a}).

\newpage

\begin{figure}[t!]
\includegraphics[width=\columnwidth]{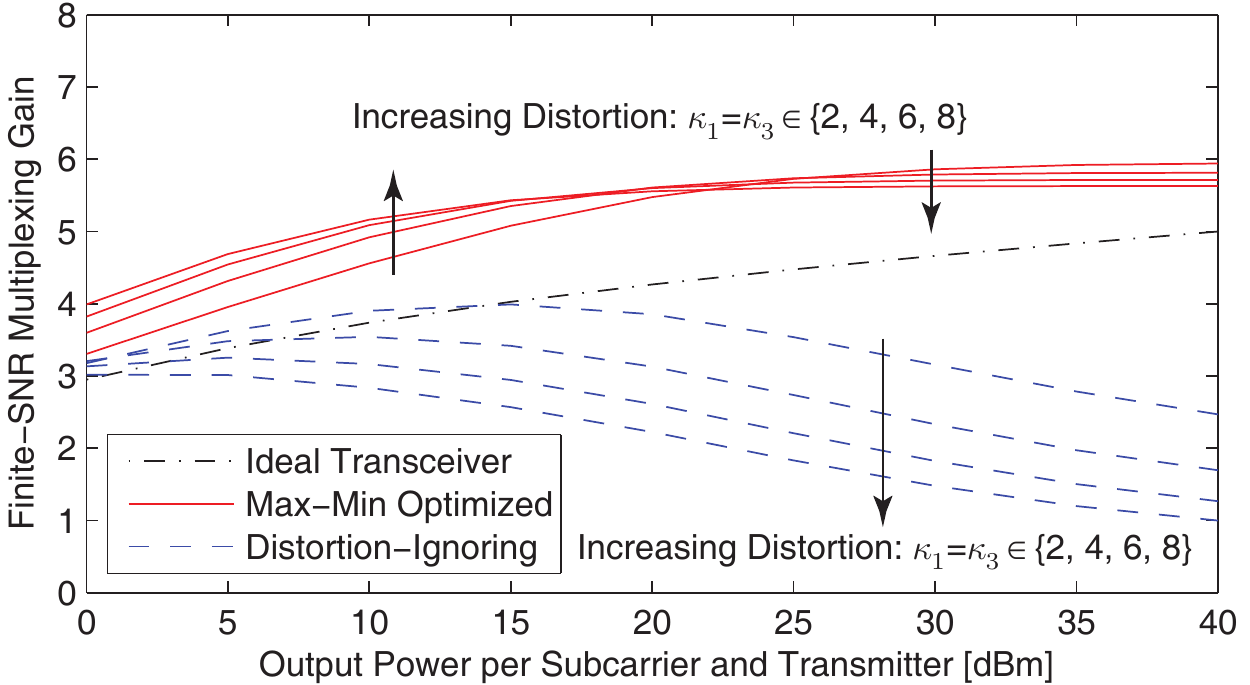} \vskip -2mm
\caption{Finite-SNR multiplexing gain as a function of the output power (for different transceiver impairments) and for the same scenario as in Fig.~\ref{figure_SNR}.}\label{figure_multiplexinggain} \vskip -6mm
\end{figure}

\enlargethispage{2mm}

\bibliographystyle{IEEEtran}
\bibliography{IEEEabrv,refs}

\begin{thebibliography}{10}
\providecommand{\url}[1]{#1}
\csname url@samestyle\endcsname
\providecommand{\newblock}{\relax}
\providecommand{\bibinfo}[2]{#2}
\providecommand{\BIBentrySTDinterwordspacing}{\spaceskip=0pt\relax}
\providecommand{\BIBentryALTinterwordstretchfactor}{4}
\providecommand{\BIBentryALTinterwordspacing}{\spaceskip=\fontdimen2\font plus
\BIBentryALTinterwordstretchfactor\fontdimen3\font minus
  \fontdimen4\font\relax}
\providecommand{\BIBforeignlanguage}[2]{{%
\expandafter\ifx\csname l@#1\endcsname\relax
\typeout{** WARNING: IEEEtran.bst: No hyphenation pattern has been}%
\typeout{** loaded for the language `#1'. Using the pattern for}%
\typeout{** the default language instead.}%
\else
\language=\csname l@#1\endcsname
\fi
#2}}
\providecommand{\BIBdecl}{\relax}
\BIBdecl

\bibitem{Gesbert2010a}
D.~Gesbert, S.~Hanly, H.~Huang, S.~Shamai, O.~Simeone, and W.~Yu, ``Multi-cell
  {MIMO} cooperative networks: A new look at interference,'' \emph{{IEEE} J.
  Sel. Areas Commun.}, vol.~28, no.~9, pp. 1380--1408, 2010.

\bibitem{Bjornson2011a}
E.~Bj{\"{o}}rnson, N.~Jald{\'e}n, M.~Bengtsson, and B.~Ottersten, ``Optimality
  properties, distributed strategies, and measurement-based evaluation of
  coordinated multicell {OFDMA} transmission,'' \emph{{IEEE} Trans. Signal
  Process.}, vol.~59, no.~12, pp. 6086--6101, 2011.

\bibitem{Dahrouj2010a}
H.~Dahrouj and W.~Yu, ``Coordinated beamforming for the multicell multi-antenna
  wireless system,'' \emph{{IEEE} Trans. Wireless Commun.}, vol.~9, no.~5, pp.
  1748--1759, 2010.

\bibitem{Bjornson2012a}
E.~Bj{\"{o}}rnson, G.~Zheng, M.~Bengtsson, and B.~Ottersten, ``Robust monotonic
  optimization framework for multicell {MISO} systems,'' \emph{{IEEE} Trans.
  Signal Process.}, vol.~60, no.~5, pp. 2508--2523, 2012.

\bibitem{Rashid1998a}
F.~Rashid-Farrokhi, K.~Liu, and L.~Tassiulas, ``Transmit beamforming and power
  control for cellular wireless systems,'' \emph{{IEEE} J. Sel. Areas Commun.},
  vol.~16, no.~8, pp. 1437--1450, 1998.

\bibitem{Bengtsson2001a}
M.~Bengtsson and B.~Ottersten, ``Optimal and suboptimal transmit beamforming,''
  in \emph{Handbook of Antennas in Wireless Communications}, L.~C. Godara,
  Ed.\hskip 1em plus 0.5em minus 0.4em\relax CRC Press, 2001.

\bibitem{Wiesel2006a}
A.~Wiesel, Y.~Eldar, and S.~Shamai, ``Linear precoding via conic optimization
  for fixed {MIMO} receivers,'' \emph{{IEEE} Trans. Signal Process.}, vol.~54,
  no.~1, pp. 161--176, 2006.

\bibitem{Tolli2009c}
A.~T\"{o}lli, H.~Pennanen, and P.~Komulainen, ``Distributed coordinated
  multi-cell transmission based on dual decomposition,'' in \emph{Proc.~IEEE
  GLOBECOM'09}, 2009.

\bibitem{Bjornson2013b}
E.~Bj{\"{o}}rnson and E.~Jorswieck, ``Optimal resource allocation in
  coordinated multi-cell systems,'' \emph{Foundations and Trends in
  Communications and Information Theory}, submitted.

\bibitem{Holma2011a}
H.~Holma and A.~Toskala, \emph{{LTE} for {UMTS}: {Evolution} to
  {LTE}-Advanced}, 2nd~ed.\hskip 1em plus 0.5em minus 0.4em\relax Wiley, 2011.

\bibitem{Dardari2000a}
D.~Dardari, V.~Tralli, and A.~Vaccari, ``A theoretical characterization of
  nonlinear distortion effects in {OFDM} systems,'' \emph{{IEEE} Trans.
  Commun.}, vol.~48, no.~10, pp. 1755--1764, 2000.

\bibitem{Studer2011a}
C.~Studer, M.~Wenk, and A.~Burg, ``System-level implications of residual
  transmit-{RF} impairments in {MIMO} systems,'' in \emph{Proc.~EUCAP'11},
  2011.

\bibitem{Studer2010a}
------, ``{MIMO} transmission with residual transmit-{RF} impairments,'' in
  \emph{Proc.~ITG/IEEE WSA'10}, 2010.

\bibitem{Galiotto2009a}
C.~Galiotto, Y.~Huang, N.~Marchetti, and M.~Zorzi, ``Performance evaluation of
  non-ideal {RF} transmitter in {LTE}/{LTE}-advanced systems,'' in
  \emph{Proc.~European Wireless}, 2009, pp. 266--270.

\bibitem{Gonzalez2011b}
J.~Gonz\'{a}lez-Coma, P.~Castro, and L.~Castedo, ``Impact of transmit
  impairments on multiuser {MIMO} non-linear transceivers,'' in \emph{Proc.~ITG
  WSA}, 2011.

\bibitem{Zetterberg2011a}
P.~Zetterberg, ``Experimental investigation of {TDD} reciprocity-based
  zero-forcing transmit precoding,'' \emph{EURASIP J. on Adv. in Signal
  Process.}, jan 2011.

\bibitem{MinicircuitsHXG-122+}
\BIBentryALTinterwordspacing
\emph{{LTE} Performance vs. Output Power, {Model}: HXG-122+}.\hskip 1em plus
  0.5em minus 0.4em\relax Mini-Circuits. [Online]. Available:
  \url{http://www.minicircuits.com/app/AN60-050.pdf}
\BIBentrySTDinterwordspacing

\bibitem{cvx}
M.~Grant and S.~Boyd, ``{CVX}: Matlab software for disciplined convex
  programming,'' \url{http://cvxr.com/cvx}, Apr. 2011.

\bibitem{LTE2010b}
\emph{Further advancements for {E-UTRA} physical layer aspects (Release
  9)}.\hskip 1em plus 0.5em minus 0.4em\relax {3GPP} {TS} 36.814, Mar. 2010.

\bibitem{Narasimhan2005b}
R.~Narasimhan, ``Finite-{SNR} diversity performance of rate-adaptive {MIMO}
  systems,'' in \emph{Proc.~IEEE GLOBECOM'05}, 2005.

\end{thebibliography}

\end{document}